\documentclass[]{JHEP3}   
 \preprint{  }

\usepackage{epsfig,multicol}
\usepackage{amsmath}
\usepackage{graphics}
\usepackage{graphicx}
\usepackage{dcolumn}
\usepackage{amssymb}

\title{A general holographic metal/superconductor phase transition model}

\author{Yan Peng$^{1}$\thanks{Electronic address: yanpengphy@163.com}, Yunqi Liu$^{2}$\thanks{Electronic address:
liuyunqi@sjtu.edu.cn}
\\$^{1}$School of Mathematics and Computer Science, Shaanxi University of Technology, Chaoyang road,  Hanzhong, Shaanxi 723000, P. R. China \\
$^{2}$Department of Physics and Astronomy, Shanghai Jiao Tong University, Dongchuan road, Shanghai 200240, P. R. China.}

\abstract{We study the scalar condensation of a general holographic superconductor model in AdS black hole background
away from the probe limit.
We find the model parameters together with the scalar mass and backreaction can determine the order of phase transitions completely.
In addition, we observe two types of discontinuities of the scalar operator in the case of first order phase transitions.
We analyze in detail the effects of the scalar mass and backreaction on the formation of discontinuities and arrive at an approximate relation between the threshold model parameters. Furthermore, we obtain superconductor solutions corresponding to higher energy states and examine the stability of these superconductor solutions.}


\begin{document}

\section{Introduction}

The AdS/CFT correspondence has provided us a useful approach to describe
strongly interacting systems holographically
through weakly coupled gravitational duals \cite{Maldacena,S.S.Gubser-1,E.Witten}.
One of the mostly studied gravity dual is the holographic superconductor,
which is constructed by a scalar field coupled to a Maxwell field in an AdS
black hole background \cite{S.A. Hartnoll,C.P. Herzog,G.T. Horowitz-1}. It shows that the black hole becomes unstable
and the scalar field condensates on the black hole background when the Hawking temperature of a black hole drops below
a critical value. According to the holographic theory, this instability in the (d+1) dimensional AdS black hole corresponds to a d dimensional metal/superconductor phase transition on the boundary. Generally speaking, this phase transition belongs to the second order.
Since then a lot of holographic duals have been
established in various gravity theories
and the
models turns out to be quite successful in giving the qualitative properties
of superconductivity. Some other recent
progress on holographic superconductors can be found in \cite{G.T. Horowitz-2}-\cite{CD}.

With a very complete holographic superconductor model in AdS black hole, it was announced in \cite{G-1} that the scalar operator as a function of the temperature is always continuous. The instability of this gravity system
corresponds to the second order phase transition.
Lately, it was stated in
\cite{S. Franco,S. Franco-1} that the holographic superconductor with the spontaneous breaking of a global U(1) symmetry via
the St\"{u}ckelberg mechanism allows the first order discontinuous phase
transition to occur. Some further studies were carried out in \cite{Q.Y.
Pan-1,YanPanCTP,P. Yan,Yan Peng-1} by considering the matter fields' backreaction on the background. It was found in \cite{P. Yan} that the light
backreaction can trigger the first order phase transition but the heavy backreaction suppresses the first order phase transition in the AdS black hole background.
Generally speaking, the phase transition is between normal state and superconducting state.
Very recently a $\psi^{2}+\zeta\psi^{6}$ St\"{u}ckelberg
mechanism was discussed in the AdS soliton spacetime \cite{cai-2},
which in particular admits new types of phase transitions between superconducting states.
So it is very interesting to extend this new St\"{u}ckelberg mechanism to the metal/superconductor system to explore the rich properties of holographic superconductors.

The entanglement entropy is usually applied to study the degrees of freedom in strongly interacting systems when other methods
might not be available.
Ryu and Takayanagi \cite{S-1,S-2} have presented a
simple and elegant way to calculate the entanglement entropy of a
strongly interacting system from a gravity dual.
The entanglement entropy has been applied to study the properties of phase transitions in various gravity theories
and proved to be a good probe to give us
new insights into the holographic superconductor models \cite{NishiokaJHEP}-\cite{W}.
In the AdS black hole background,
Albash and Johnson observed in \cite{T-6} that
the entanglement entropy as a function of
temperature has a discontinuous slop at the transition
temperature $T_{c}$ corresponding to the second order metal/superconductor phase transition. In contrast, there is a jump in the entanglement entropy
when allowing the first order phase transition \cite{T-6,Yan Peng-1,Cai-4,Cai-5}, which
means that the entanglement entropy can be used to disclose
the order of phase transitions.
When considering a new St$\ddot{u}$ckelberg model with a $\psi^{2}+\zeta\psi^{6}$ term in the AdS soliton background \cite{cai-2},
the solutions in particular admits richer structures for the insulator/superconductor phase transition diagram.
As a further step along this line,
it is of great interest to generalize the investigation in \cite{cai-2} to AdS black hole
and study general features of the metal/superconductor phase transitions through entanglement entropy approach
in such a $\psi^{2}+\zeta\psi^{6}$ St$\ddot{u}$ckelberg holographic superconductor model.

At zero temperature, it was found in \cite{RA} that different energy states appear for the scalar condensation
in the AdS black hole background. Superconductor solutions corresponding to different states were also mentioned with the behaviors of the scalar
fields away from the zero temperature limit \cite{GT}.
When applying the discussion to holographic insulator/superconductor system, different states will turn out by increasing the critical chemical potential from zero to
critical values with other parameters fixed \cite{Cai-7}. It concluded in \cite{Cai-7} that the second and third states are less stable
due to the oscillations of scalar field in the radial direction and the first state is related to the superconductor solutions in the AdS soliton spacetime \cite{T-1}.
It is interesting to extend the discussion to examine whether there are different states in more general holographic superconductor models and also further explore the stability of various possible states.

The next section is organized as follows. In part A, we introduce
a St$\ddot{u}$ckelberg holographic
model in the four dimensional AdS black hole background. And in part B,
we discuss in detail the scalar condensation and the holographic entanglement entropy
of the system in superconductor phase. Part C is devoted to the stability of various solutions with different energy states.
We summarize
and discuss our main results in section III.

\section{a general model of superconductor in AdS black hole}

\subsection{Equations of motion and boundary conditions}

We begin with a generalized St\"{u}ckelberg Lagrange density containing a Maxwell field and a scalar field,
\begin{eqnarray}\label{lagrange-1}
\mathcal{L}=R+\frac{6}{L^{2}}-\gamma[\frac{1}{4}F^{\mu\nu}F_{\mu\nu}+(\partial
\psi)^{2}+m^{2}|\psi|^{2}+G(\psi)(\partial p-A)^{2}],
\end{eqnarray}
where $A_{\mu}$ and $\psi$ are the Maxwell field and a charged scalar field with mass $m^{2}$, respectively. $-3/L^{2}$ is the negative cosmological constant, where $L$ is the AdS radius
which will be scaled unity in our calculation. $\gamma$ is the backreaction parameter describing the effects of matter fields on the background. When $\gamma\rightarrow 0$,
the backreaction of the matter fields on the background becomes negligible and the
metric solutions reduce to the pure AdS black hole spacetime. We
will establish a general St\"{u}ckelberg holographic model in AdS black hole by considering a simple form $G(\psi)=\psi^{2}+\zeta\psi^{n}$
with $n\in N$, where
$\zeta$ is the model parameter.
With the gauge symmetries
$A\rightarrow A+\partial\Lambda$ and $p\rightarrow
p+\Lambda$, we fix $p=0$ in the following discussion.

The metric and other fields of interest are parameterized as follows:
\begin{eqnarray}\label{AdSBH}
ds^{2}&=&-g(r)e^{-\chi(r)}dt^{2}+\frac{dr^{2}}{g(r)}+r^{2}(dx^{2}+dy^{2}),\nonumber\\
A&=&\phi(r)dt,\\
\psi&=&\psi(r).\nonumber
\end{eqnarray}
With this ansatz, the Hawking temperature of the black hole is given by
\begin{eqnarray}\label{HawkingT}
T_{H}=\frac{g'(r_{+})e^{-\chi(r_{+})/2}}{4\pi},
\end{eqnarray}
where $r_{+}$ corresponds to the horizon of the black hole satisfying $g(r_{+})=0$.

We can obtain equations of motion from the action
\begin{eqnarray}\label{BHChi}
\chi'+\gamma\left[r\psi'^{2}+\frac{r}{g^{2}}e^{\chi}\phi^{2}G(\psi)\right]=0,
\end{eqnarray}
\begin{eqnarray}\label{BHg}
g'-\left(\frac{3r}{L^{2}}-\frac{g}{r}\right)+
\gamma rg\left[\frac{1}{2}\psi'^{2}+\frac{1}{4g}e^{\chi}\phi'^{2}+\frac{m^{2}}{2g}\psi^{2}+\frac{1}{2g^{2}}e^{\chi}\phi^{2}G(\psi)\right]=0,
\end{eqnarray}
\begin{eqnarray}\label{BHphi}
\phi''+\left(\frac{2}{r}+\frac{\chi'}{2}\right)\phi'-\frac{2G(\psi)}{g}\phi=0,
\end{eqnarray}
\begin{eqnarray}\label{BHpsi1}
\psi''+\left(\frac{2}{r}-\frac{\chi'}{2}+\frac{g'}{g}\right)\psi'-\frac{m^{2}}{g}\psi+\frac{1}{2g^{2}}e^{\chi}\phi^{2}G'(\psi)=0,
\end{eqnarray}
where $G'(\psi)$ represents the derivative with respect to $\psi$.
Since the equations are coupled and nonlinear, we have to solve these equations
by numerically integrating them from the horizon out to the
infinity.

By considering behaviors of solutions at the horizon $r_{+}$, we find that there are four independent parameters $r_{+}$,
$\psi(r_{+})$, $\phi'(r_{+})$ and $\chi(r_{+})$ at the horizon. The scaling symmetry
\begin{eqnarray}\label{symmetryBH1}
r \rightarrow ar,~~~~~~~~t\rightarrow
~at,~~~~~~~\phi\rightarrow
a\phi,~~~~~~g\rightarrow\ a^{2} g,
\end{eqnarray}
can be used to set $r_{+}=1$. These equations are
also invariant under another scaling,
\begin{eqnarray}\label{symmetryBH2}
e^{-\chi} \rightarrow b^{2}e^{-\chi},~~~~~~~~t\rightarrow
bt,~~~~~~~\phi\rightarrow
\phi/b,
\end{eqnarray}
which enable us to choose an arbitrary value of $\chi(r_{+})$.  With this Eq.(\ref{symmetryBH}), we set $\chi(r\rightarrow\infty)=0$ to recover the AdS boundary.

Near the asymptotic
AdS boundary $(r\rightarrow \infty)$, the asymptotic behaviors of the scalar and Maxwell fields are
\begin{eqnarray}\label{InfBH}
\psi=\frac{\psi_{-}}{r^{\lambda_{-}}}+\frac{\psi_{+}}{r^{\lambda_{+}}}+\cdot\cdot\cdot,\
\phi=\mu-\frac{\rho}{r}+\cdot\cdot\cdot, \ \
\end{eqnarray}
with $\lambda_{\pm}=(3\pm\sqrt{9+4m^{2}})/2$, where $\mu$ and $\rho$ can be interpreted as
the chemical potential and charge density in the dual theory
respectively.
When $\psi(r)=0$, we get the analytic solutions in normal phase,
a Reissner-Nordstrom-AdS black hole, which is given by
\begin{eqnarray}\label{}
g=\frac{r^{2}}{L^{2}}-\frac{2M}{r}+\frac{\gamma\rho^{2}}{4r^{2}} ,~~\chi=0,~~~\phi=\rho(\frac{1}{r_{+}}-\frac{1}{r}),
\end{eqnarray}
where M is the integration constant that can be interpreted as the mass of the
black hole.
It is known that the black hole is unstable when the temperature T is smaller than a critical temperature $T_{c}$. For low
temperature $T<T_{c}$, a hairy black hole with $\psi(\infty)=0$ appears.
To get the hairy black hole, we will fix $\psi_{-}=0$ and the phase transition in the dual CFT is described by the
operator $\psi_{+}=<O_{+}>$ in the following discussion. For $m^{2}> -\frac{9}{4}$ above the BF bound \cite{P. Breitenlohner}, the second scalar operator $\psi_{+}$ is always normalizable.
For each fixed value $\psi(r_{+})$ at the horizon, we take $\phi'(r_{+})$ as the shooting parameter to search for the solutions satisfying
the boundary conditions $\psi_{-}=0$.

\subsection{The scalar condensation in AdS black hole}

It was stated that a $\psi^{2}+\zeta\psi^{6}$ St\"{u}ckelberg
mechanism brings richer physics in the holographic insulator/superconductor phase transitions in the AdS soliton spacetime \cite{cai-2}.
In the holographic metal/superconductor phase transition, we find the results are qualitative similar when $n\geqslant5$.
So we will focus on the case of $n=6$ in this paper similar to the discussions in Ref. \cite{cai-2} for simplicity.
We start with studying
the free energy of the system. The free energy of the field theory is determined by the integral of the lagrange density Eq.(1) evaluated on-shell or $F=-T\widetilde{S}$, where
$\widetilde{S}=\frac{1}{V}\int LdV$.
However, this integration will in general be divergent and suitable
counterterms have to be added \cite{Sean A. Hartnoll-3,Y. Brihaye-1}. The interesting
quantity is the difference in free energy between the superconductor and normal phases.

\begin{eqnarray}\label{lagrange-1}
\Delta F=F_{superconductor}-F_{normal}.
\end{eqnarray}

We show the free energy as a function of temperature in Fig. 1 with $\gamma=0.1$, $m^{2}=-2$ and various $\zeta$.
It can be seen from $(a),~(b)$ and $(c)$ in Fig. 1 that, for the small model parameters ($\zeta=0,~0.2 ~or~ 0.3$), $\Delta F$
decreases smoothly near the critical temperature $T_c$ indicating the second order phase
transitions from normal state into superconducting state. What's more,
in $(c)$ the plot with $\zeta=0.3$, besides the second order phase transition at the critical temperature $T_c$,
the free energy develops a ``swallow tail'' at $T\thickapprox 0.0506$, a typical signal for a first order
phase transition, indicating that there is a new phase transition within the superconducting phase.
$(d)$ of Fig. 1 shows that as $\zeta$ increases to 0.6, $\Delta F$ develops a discontinuity in the first derivative of the free energy with respect to temperature at critical temperature $T_c$, which implies that a strong St$\ddot{u}$ckelberg mechanism with $\zeta=0.6$ triggers the first order phase transition.

\begin{figure}[h]
\centering
\includegraphics[width=180pt]{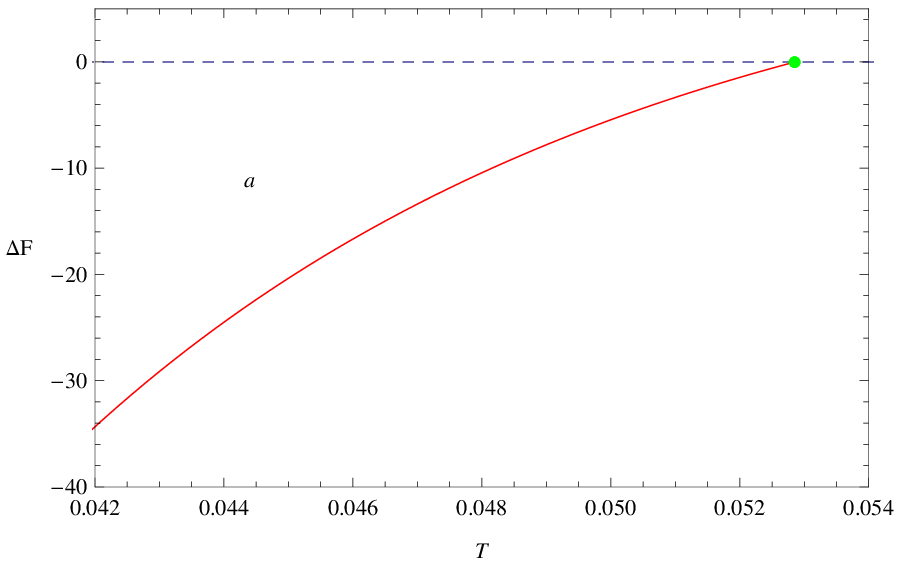}\
\includegraphics[width=180pt]{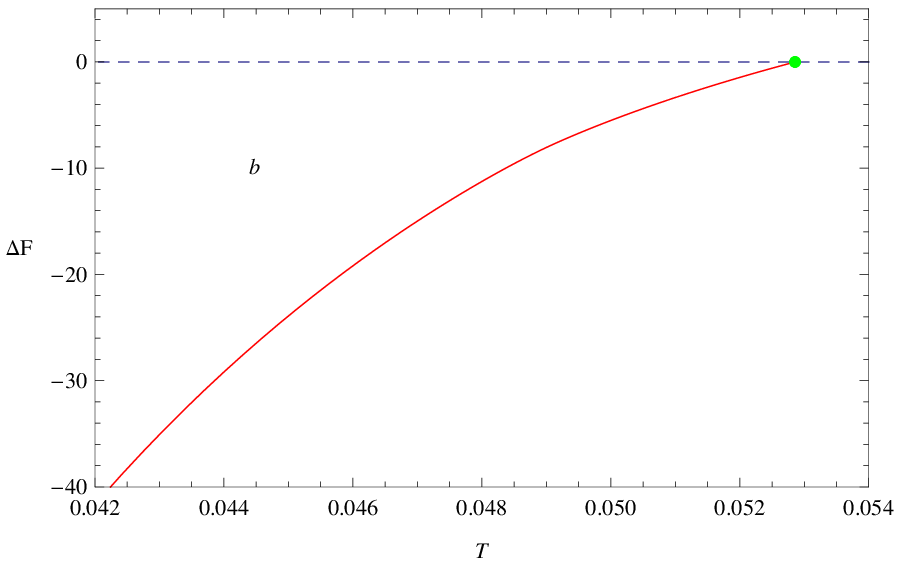}\
\includegraphics[width=180pt]{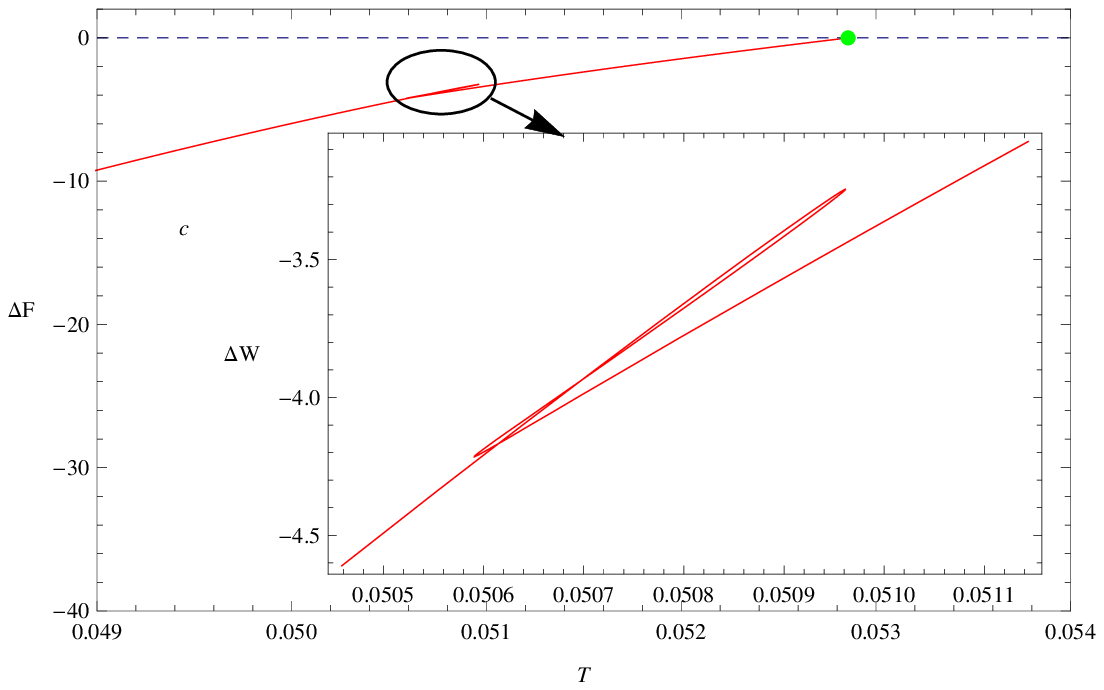}\
\includegraphics[width=180pt]{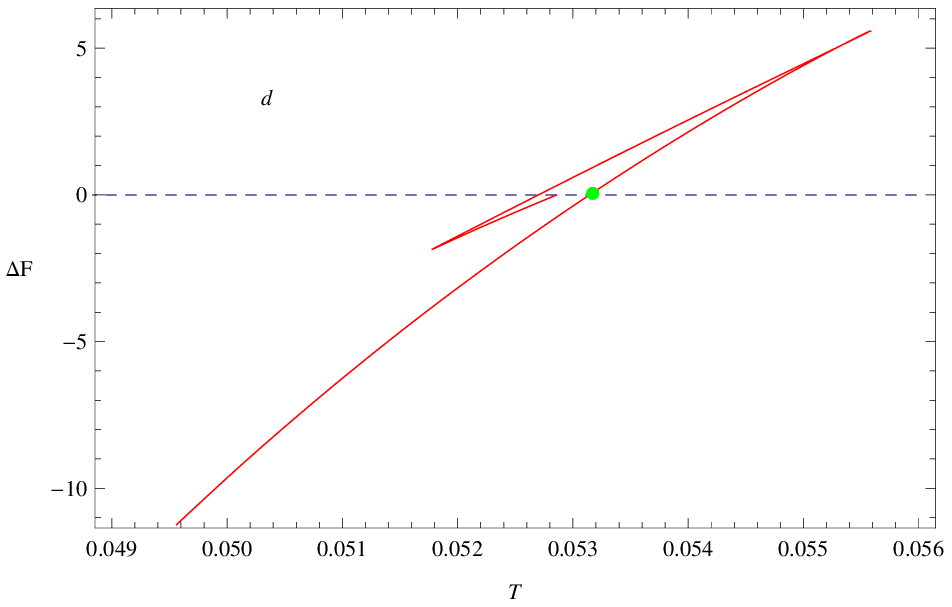}\
\caption{\label{EEntropySoliton} (Color online) The free energy as
a function of $T$ with $\mu=1$, $m^{2}=-2$, $\gamma=0.1$ and various $\zeta$: $(a)$ the case $\zeta=0$,
$(b)$ the case $\zeta=0.2$, $(c)$ the case $\zeta=0.3$, $(d)$ the case $\zeta=0.6$. The blue dashed line in each panel corresponds to the free energy of
a black hole in normal state. And the green solid points in each panel indicate the critical phase transition point $T_{c}$ between normal and superconductor phase.}
\end{figure}

We could also detect the properties of phase transitions by studying the condensation of the scalar operator
for different values of $\zeta$.
In Fig. 2, we can see that the operator
$\langle O_{+}\rangle^{1/\lambda_{+}}$ is monotonically as a function of temperature in $(a)$ with $\zeta=0$ and $(b)$ with $\zeta=0.2$ around the phase transition point.
However, in $(c)$ as we choose the parameter $\zeta=0.3$,
the curve starts from $\langle O_{+}\rangle^{1/\lambda_{+}}=0$ at the critical temperature $T_{c}$ signaling a second order phase transition, and a dump of the scalar operator appears at $T\thickapprox 0.0506$ corresponding to the first order discontinuity in $(c)$ of Fig. 1.
In cases that the transition occurs at $T_{c}$, the normal phase transfers into superconducting phase as we decrease
the temperature. When the phase transition is second order, we have checked the mean field exponents and find the
condensate approaches zero as $<O_{+}>\varpropto(T_{c}-T)^{1/2}$.
This is reasonable since $G(\psi)=\psi^{2}+\zeta \psi^{6}\thickapprox\psi^{2}$ (or independent of $\zeta$) around the phase transition point $T_{c}$.
For example, the curves in (a), (b) and (c) of Fig. 2 correspond to $<O_{+}>\thickapprox2.4(T_{c}-T)^{1/2}$ around $T_{c}$.
As the model parameter $\zeta$ grows up, it can be seen in (d) of Fig. 2. that the operator $\langle O_{+}\rangle^{1/\lambda_{+}}$ starts from a finite value $0.49$ at $T_c$, which means the transition from the normal phase to the superconductor belongs to the first order phase transition.

\begin{figure}[h]
\centering
\includegraphics[width=193pt]{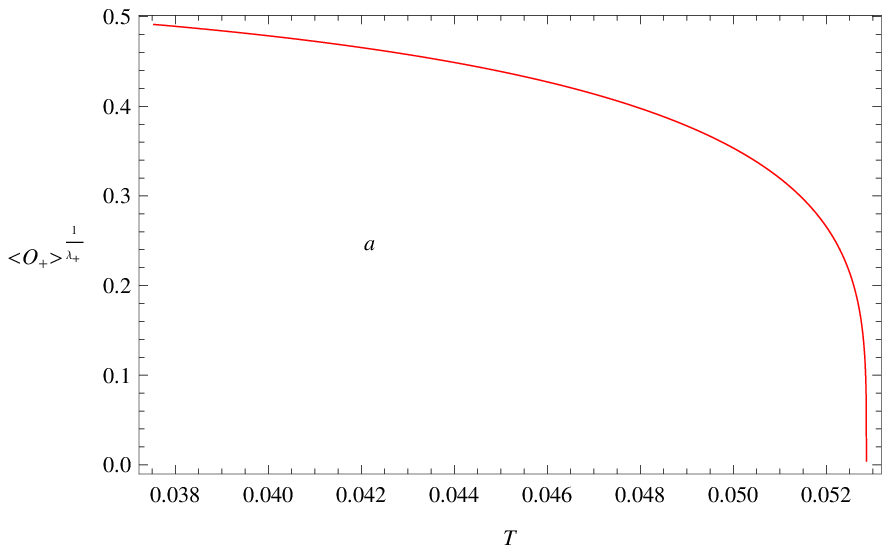}\
\includegraphics[width=193pt]{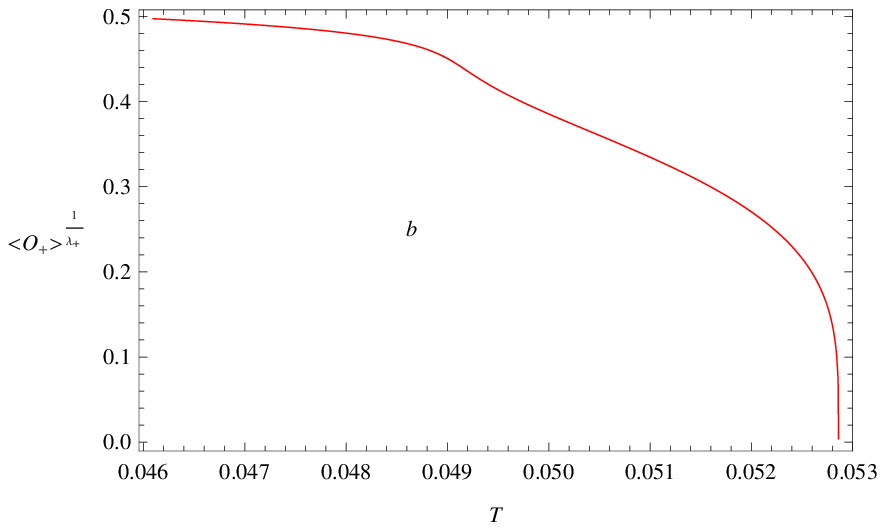}\
\includegraphics[width=193pt]{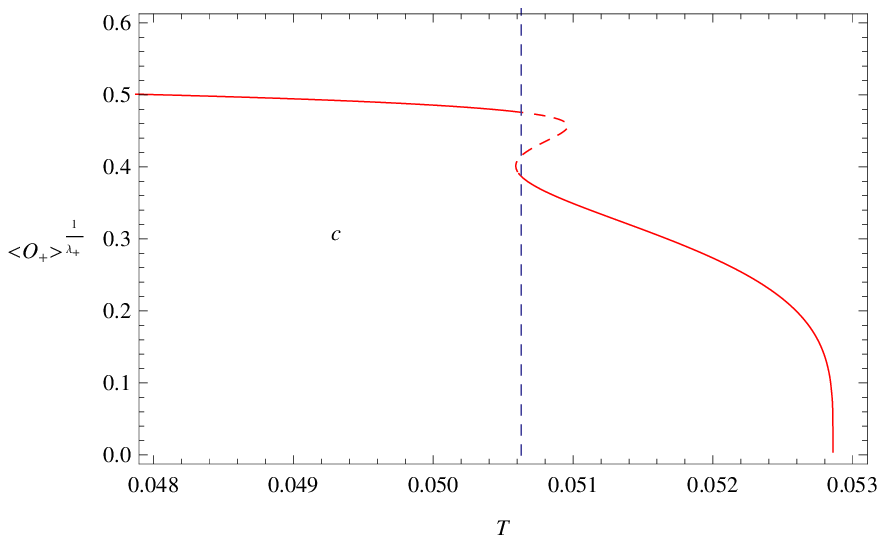}\
\includegraphics[width=193pt]{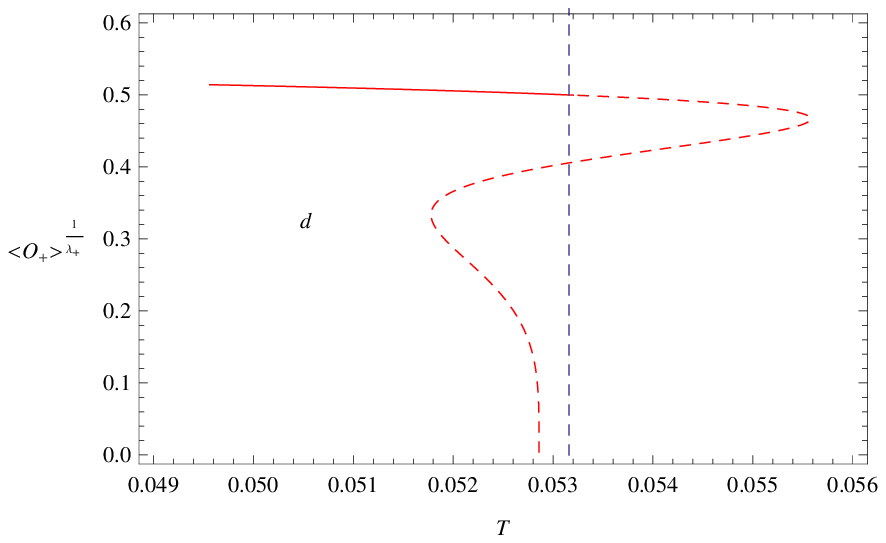}\
\caption{\label{EEntropySoliton} (Color online) The scalar operator as
a function of $T$. We take $\mu=1$, $m^{2}=-2$, $\gamma=0.1$ and $\zeta$ varies as:
$(a)$ the case $\zeta=0$,
$(b)$ the case $\zeta=0.2$, $(c)$ the case $\zeta=0.3$, $(d)$ the case $\zeta=0.6$.
The red solid line in each panel corresponds to the superconductor phase.
}
\end{figure}

For the rich phases structure in the boundary theory in Fig. 1, in the gravity side we can use the gravity solotions
to characterize the difference between these phases. In Fig. 3, fixing the coordinate $r=r_{0}=5$ (outside the horizon or $r_{0}\geqslant r_{+}$), we depict the metric
$g(r_{0})$ as a function of temperature T.
In (a), (b) and (c) of Fig. 3,
we find $g(r_{0})$ has a discontinuous slop with respect to T at the temperature
 $T=0.05286$, which corresponding to the second order phase transition points in (a), (b) and (c) of Fig. 1.
 Further, in the first order phase transition points in (c) and (d) between superconducting phases in Fig. 1,
 we find jumps of $g(r_{0})$.
Then we concluded that the discontinuity in the first derivative of $g(r_{0})$ corresponds to the second order phase transition
and the jump of $g(r_{0})$ corresponds to the first order phase transition.

\begin{figure}[h]
\centering
\includegraphics[width=180pt]{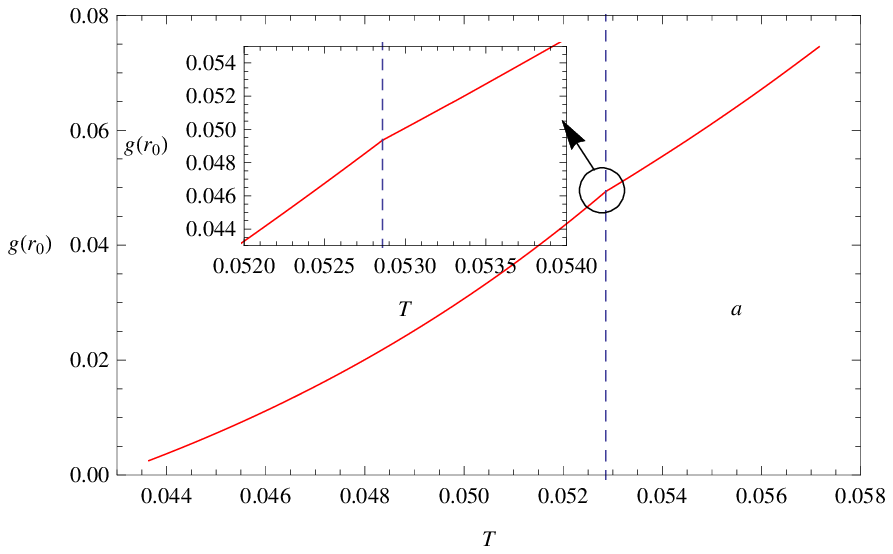}\
\includegraphics[width=180pt]{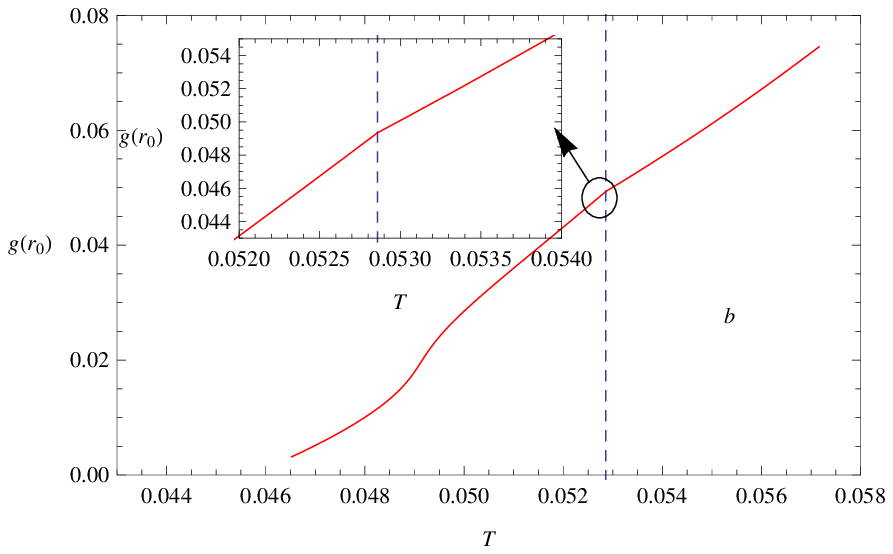}\
\includegraphics[width=180pt]{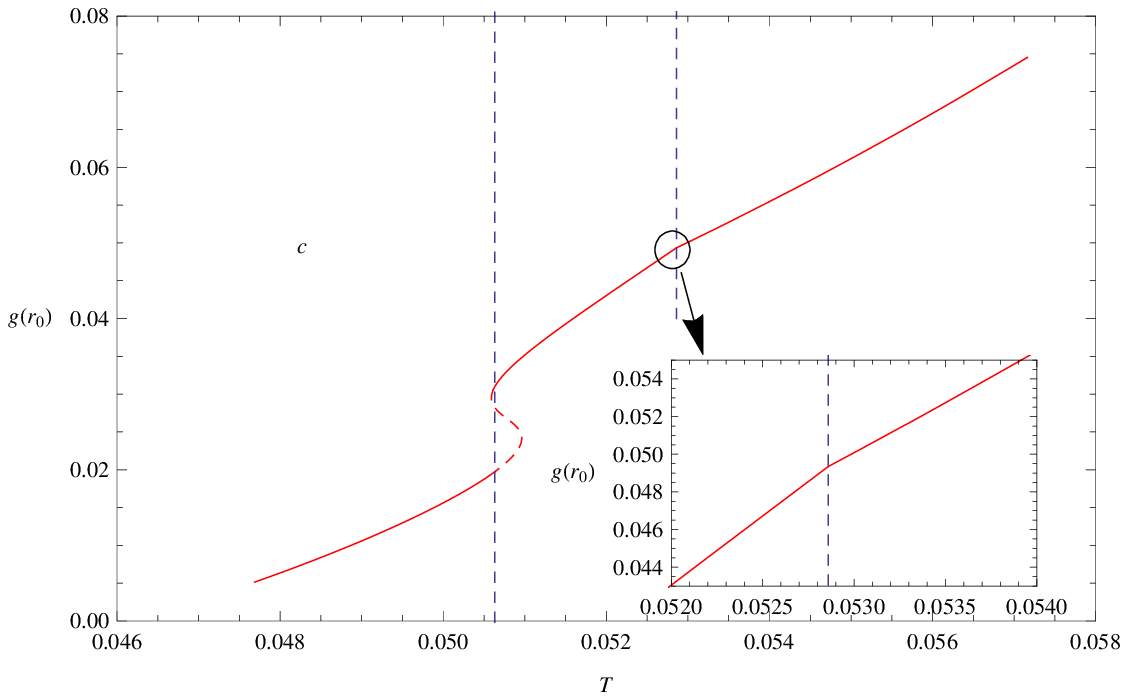}\
\includegraphics[width=180pt]{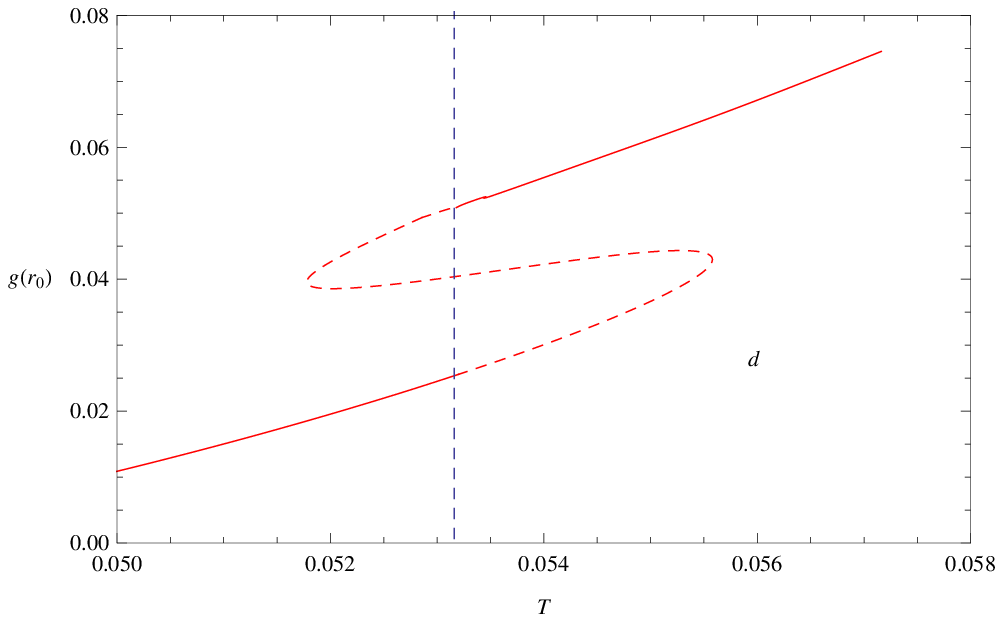}\
\caption{\label{EEntropySoliton} (Color online) The metric solution $g(r_{0})$ as
a function of $T$ with $r_{0}=5$, $\mu=1$, $m^{2}=-2$, $\gamma=0.1$ and various $\zeta$: $(a)$ the case $\zeta=0$,
$(b)$ the case $\zeta=0.2$, $(c)$ the case $\zeta=0.3$, $(d)$ the case $\zeta=0.6$.
}
\end{figure}

In the following discussion, we pay our attention to the holographic entanglement entropy(HEE) of the phase transition system. The
authors in Refs. \cite{S-1,S-2} have presented a proposal to compute
the entanglement entropy of conformal field theories (CFTs) from the
minimal area surface in gravity side.
We consider a subsystem $\tilde{A}$ with a straight strip geometry
described by $-\frac{l}{2}\leqslant x\leqslant\frac{l}{2},~0\leq y \leq \tilde{L}$,
where $l$ is defined as the size of region $\tilde{A}$, and
$\tilde{L}$ is a regulator which is set to be infinity. Minimizing
the area of hypersurface $\gamma_{\tilde{A}}$ whose boundary is the
same as the stripe $\tilde{A}$, the entanglement entropy for a belt
geometry can be expressed as \cite{T-6}
\begin{eqnarray}\label{EEntropyBH}
S=\int^{z_{*}}_{\varepsilon}dz\frac{z_{*}^{2}}{z^{2}}\frac{1}{\sqrt{(z^{4}_{*}-z^{4})z^{2}g(z)}}-\frac{1}{\varepsilon},
\end{eqnarray}
with
\begin{eqnarray}\label{Length}
\frac{l}{2}=\int^{z_{*}}_{\varepsilon}dz\frac{z^{2}}{\sqrt{(z^{4}_{*}-z^{4})z^{2}g(z)}},
\end{eqnarray}
where $z_{*}$ satisfies the condition $\frac{dz}{dx}|_{z_{*}}=0$
with $z=\frac{1}{r}$ and the UV cutoff $r=\frac{1}{\varepsilon}$ has been taken into consideration.

Now we exhibit properties of the phase transition through behaviors of the entanglement entropy.
We plot the entanglement entropy as a function of the temperature $T$ in Fig. 4 with $\mu=1$, $\gamma=0.1$ and various $\zeta$.
The blue dashed lines describe the HEE of the normal phase as a function of temperature, while the red solid lines correspond to the HEE of the superconducting phase. In $(a)$ with $\zeta=0$ and $(b)$ with $\zeta=0.2$ of Fig. 4, the holographic entanglement entropy decreases continuously as temperature decreases, and there are discontinuous slops at the transition temperature $T_{c}$, which imply the phase transitions at the critical temperature $T_c$ are second order. In $(c)$ with $\zeta=0.3$, the entanglement entropy continuously decreases at phase transition point $T_c$ indicating a second order phase transition, and it also has a jump around $T=0.0506$ which corresponds to the ``swallow tail'' in free energy $\Delta F$ in $(c)$ of Fig. 1 and the dump in the order operator in $(c)$ of Fig. 2, which implies a first order phase transition in the superconducting phase. When we set the parameter $\zeta=0.6$ in $(d)$, we can see that at the critical temperature $T_c$ the entanglement entropy develops a discontinuous jump implying a first order phase transition. From above discussion, we note that the St$\ddot{u}$ckelberg mechanism provides richer physics in metal/superconductor phase transitions. The entanglement entropy can be used to distinguish the order of phase transitions in our general holographic superconductor model.

\begin{figure}[h]
\centering
\includegraphics[width=180pt]{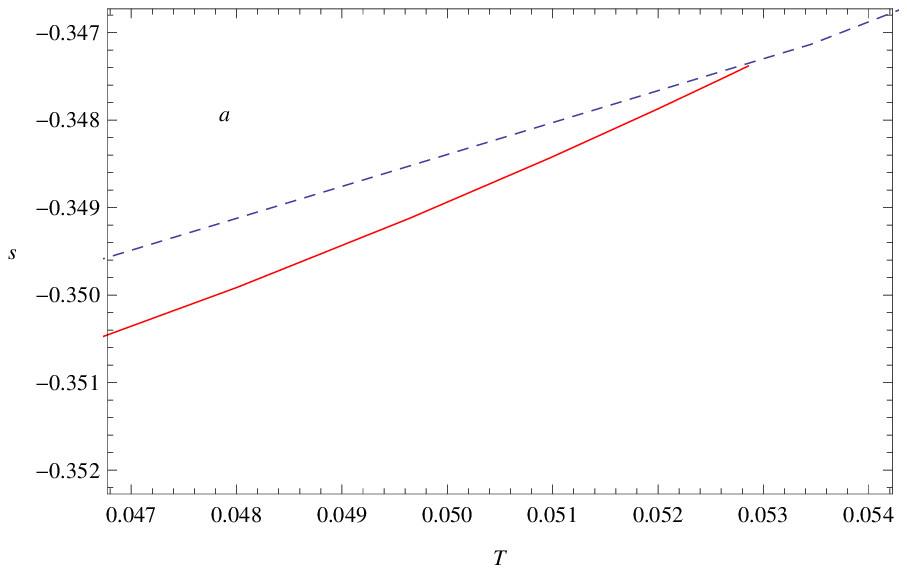}\
\includegraphics[width=180pt]{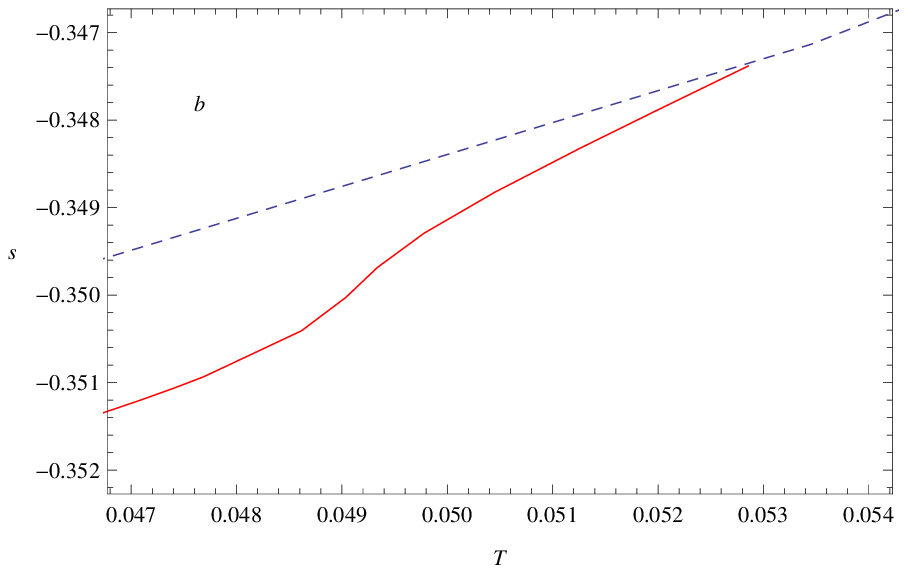}\
\includegraphics[width=180pt]{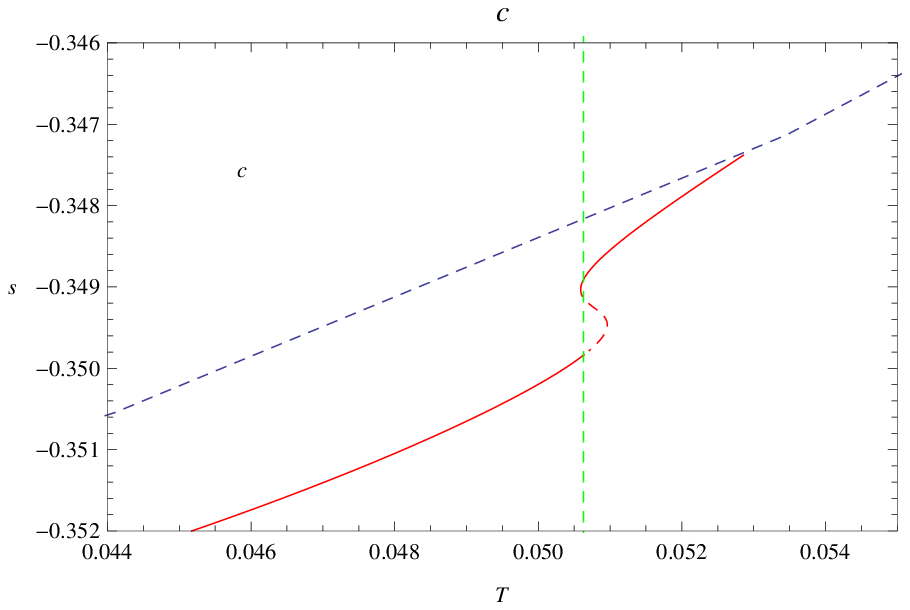}\
\includegraphics[width=180pt]{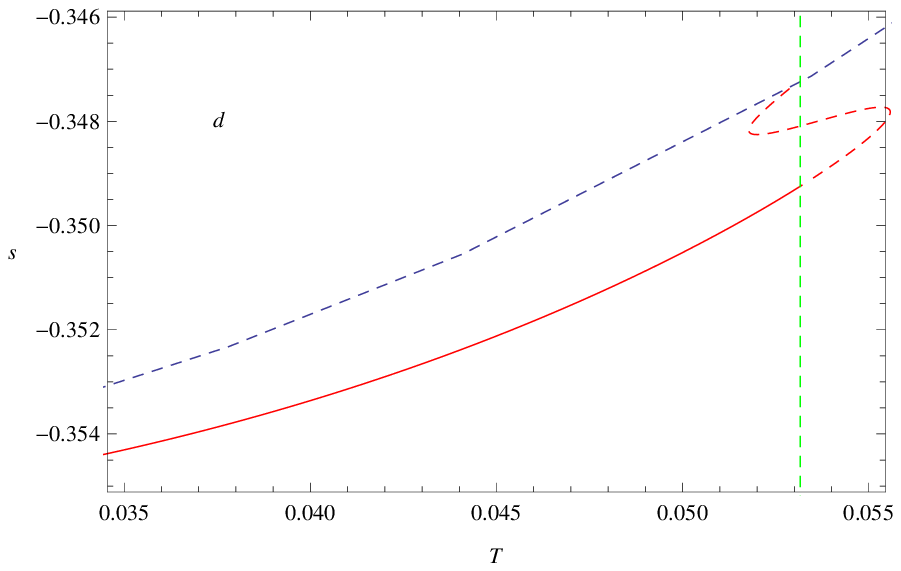}\
\caption{\label{EEntropySoliton} (Color online) The entanglement
entropy as a function of the temperature $T$ for
$\mu=1$ and $l=2$. The blue dashed line in each panel corresponds to the entanglement
entropy of a pure AdS black hole. The red solid curves show the entanglement
entropy of the superconductor phase. The cases are: (a) $\zeta=0$, $m^{2}=-2$, $\gamma=0.1$, (b) $\zeta=0.2$, $m^{2}=-2$, $\gamma=0.1$, (c)$\zeta=0.3$, $m^{2}=-2$, $\gamma=0.1$ and (d) $\zeta=0.6$, $m^{2}=-2$, $\gamma=0.1$.}
\end{figure}

According to the above discussions, we conclude that the $St\ddot{u}ckelberg$ mechanism triggers the second order phase transition at $T_{c}$
for small $\zeta$. When we choose a larger $\zeta$, an additional first order phase transition may appear in the superconducting phase,
whereas the phase transition at $T_{c}$ still corresponds to the second order. However, there is only the first order phase transition
at $T_{c}$ for very large $\zeta$.
We define two parameters $\bar{\zeta}$ and $\widetilde{\zeta}$ as threshold values,
for $\zeta\leqslant\tilde{\zeta}$ the second order phase transition appears at the critical temperature $T_c$.
And for $\tilde{\zeta} <\zeta<\bar{\zeta}$, the system experiences a second order phase transition at $T_c$ and an additional first order transition
in the superconducting phase.
When $\zeta\geqslant\bar{\zeta}$, there is only the first order phase transition at the critical temperature.
Table I shows the values of $\tilde{\zeta}$ and $\bar{\zeta}$ for various $m^{2}$ and $\gamma$.
From the table, we obtain an approximate relation
$\bar{\zeta}\thickapprox 2\tilde{\zeta}$. In order to see the effects of the backreaction $\gamma$ and mass $m^{2}$ on the condensation
more clearly, we show $\tilde{\zeta}$ and $\bar{\zeta}$ as a function of $m^{2}$ with $\gamma=0.1$ in the left panel of Fig. 5 and reach the conclusion that a less negative mass makes the first order phase transition easier to happen (or smaller threshold parameters $\tilde{\zeta}$ and $\bar{\zeta}$). With a fixed $m^{2}$ and various $\gamma$, the right panel of Fig. 5 signals that strong backreaction  promote the appearance of first order phase transitions. From the pictures, we also mention that the relation $\bar{\zeta}\thickapprox 2\tilde{\zeta}$ holds very well.

\begin{figure}[h]
\centering
\includegraphics[width=180pt]{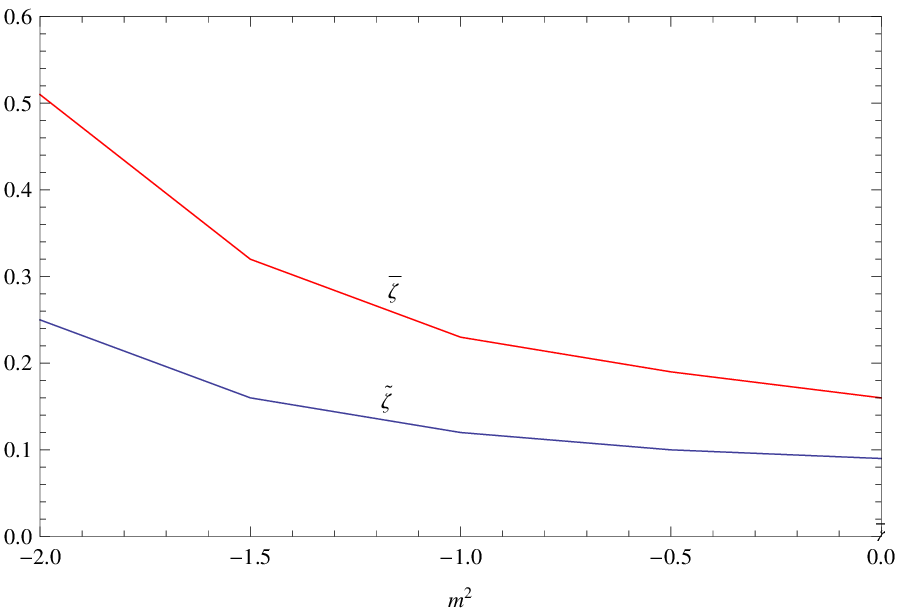}\
\includegraphics[width=180pt]{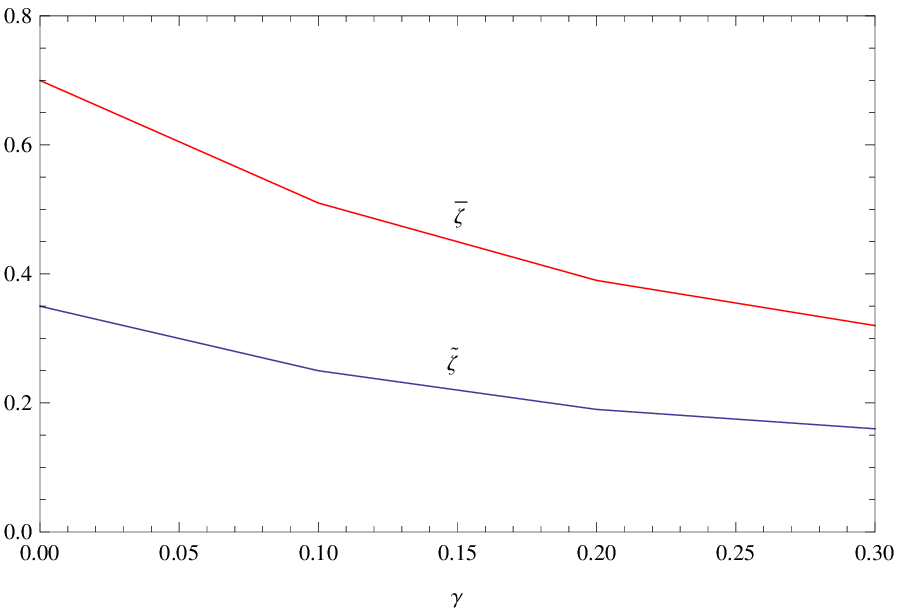}\
\caption{\label{EEntropySoliton} (Color online) The behaviors of the threshold values of $\overline{\zeta}$ (red) and  $\widetilde{\zeta}$ (blue). The left panel corresponds to the case of $\gamma=0.1$ and  various $m^{2}$. The right one shows the case $m^{2}=-2$ and  various $\gamma$.}
\end{figure}

\begin{table} \caption {The values of $\tilde{\zeta}$ and $\bar{\zeta}$ for various $m^{2}$ and $\gamma$.}
\centering
\begin{tabular}{c c c c}
         \hline
~ & $\gamma=0.1,m^{2}=-2$~~~~&~~~~
$\gamma=0.2,m^{2}=-2$~~~~ &~~~~ $\gamma=0.1,m^{2}=-1$
        \\
        \hline
        \\
~~~~$\widetilde{\zeta}$~~~~~~~~&~~~~~~~~$0.25$~~~~~~~~&~~~~~~~~$0.19$~~~~~~~~&~~~~~~~~$0.12$~~~~~~~~
          \\

          \\
~~~~$\overline{\zeta}$~~~~~~~~&~~~~~~~~$0.51$~~~~~~~~&~~~~~~~~$0.39$~~~~~~~~&~~~~~~~~$0.23$~~~~~~~~
          \\
        \hline
\end{tabular}
\end{table}

We would like to give a glance at the effects of the scalar mass on the phase
transition from the equations of motion. Let us rewrite Eq.(7) into the following form:
\begin{eqnarray}\label{BHpsi2}
\psi''+\left(\frac{2}{r}-\frac{\chi'}{2}+\frac{g'}{g}\right)\psi'-[m^{2}-\frac{1}{g}e^{\chi}\phi^{2}\left(1+3\zeta\psi^{4}\right)]\frac{\psi}{g}=0.
\end{eqnarray}
We consider the last term in the left side as the effective mass $m^{2}_{eff}(\psi)=m^{2}-\frac{1}{g}e^{\chi}\phi^{2}\left(1+3\zeta\psi^{4}\right)$.
For $\zeta=0$, it returns to the model without St\"{u}ckelberg mechanism. For $\zeta$ above the threshold parameters, this term deforms the scalar field to prompt the appearance of first
order phase transition.
When we choose a
more negative scalar mass $m^{2}$, the effective mass is mostly dominated by the scalar mass. For a less negative mass $m^{2}\approx0$, the parameter $\zeta$
will play a dominant role in the scalar condensation and we need a smaller parameters $\tilde{\zeta}$ and $\bar{\zeta}$ to trigger the first order phase transitions.

\subsection{The stability of various solutions}

For each value $\psi(r_{+})$, we find discrete values $\phi'(r_{+})$ satisfying the boundary conditions $\psi_{-}=0$.
As we choose various $\psi(r_{+})$, we obtained different families of solutions for $\psi(r)$ that satisfying the asymptotic boundary
condition. The solutions can be labeled by the number of times that $\psi(r)$ vanishes.
The plots in Fig. 6 show the scalar fields corresponding to the first three states with $\gamma=0.1, m^{2}=-2,\zeta=0$ and $\psi(r_{+})=0.1$.
The scalar field of the first state in Fig. 6 starts from $\psi(1)=0.1$ at the horizon and decreases monotonically to zero as approaching the boundary.
And the higher states of the green and blue curves correspond to scalar fields with oscillations, which are similar to the case in AdS soliton background \cite{Cai-7}.

\begin{figure}[h]
\centering
\includegraphics[width=220pt]{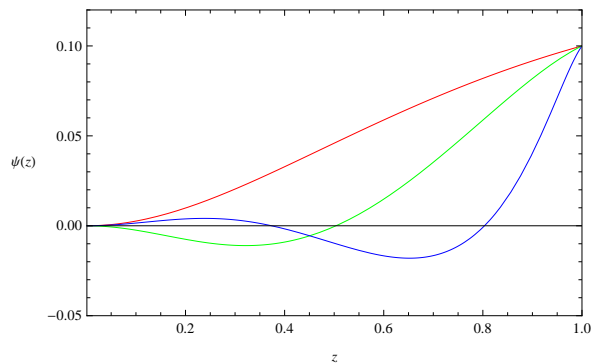}\
\caption{\label{EEntropySoliton} (Color online) The behaviors of the scalar fields $\psi(z)$ with $z=\frac{1}{r}$ and $\psi(1)=0.1$. The three curves correspond to the solutions:
the 1st state $\phi'(1)=3.929$ (Red), the 2nd state $\phi'(1)=7.809$ (Green) and the 3rd state $\phi'(1)=9.916$ (Blue) with $\gamma=0.1$, $m^{2}=-2$ and $\zeta=0$.}
\end{figure}

Now we turn to study behaviors of the scalar operator with $\gamma=0.1$, $m^{2}=-2$ and various model parameters $\zeta$ in Fig. 7. We find the St\"{u}ckelberg mechanism
can trigger first order discontinuities in scalar condensation of all states. For the second state in the left panel of Fig. 7 is with $0.01<\tilde{\zeta}<0.02$ and the third state in the right panel of Fig. 7,
 $0.003<\tilde{\zeta}<0.004$. Compared with the results $\widetilde{\zeta}=0.25$ of the first state in Table I, we find the solutions with a lower state
corresponds to a larger threshold model parameter $\widetilde{\zeta}$, above which first order phase transitions will appear in the superconducting phase.
That is to say the superconductor solutions with higher states changes dramaticly with a small perturbation of the model parameter $\zeta$.

\begin{figure}[h]
\centering
\includegraphics[width=193pt]{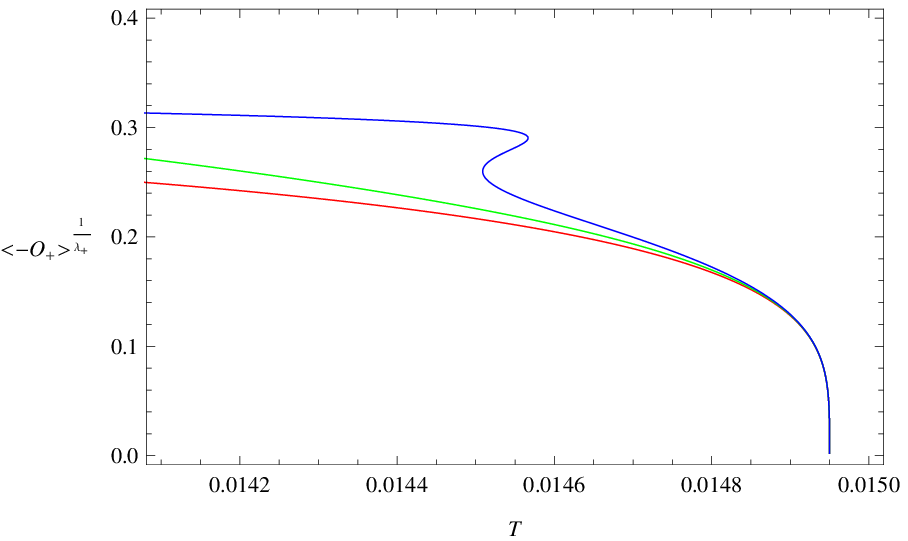}\
\includegraphics[width=193pt]{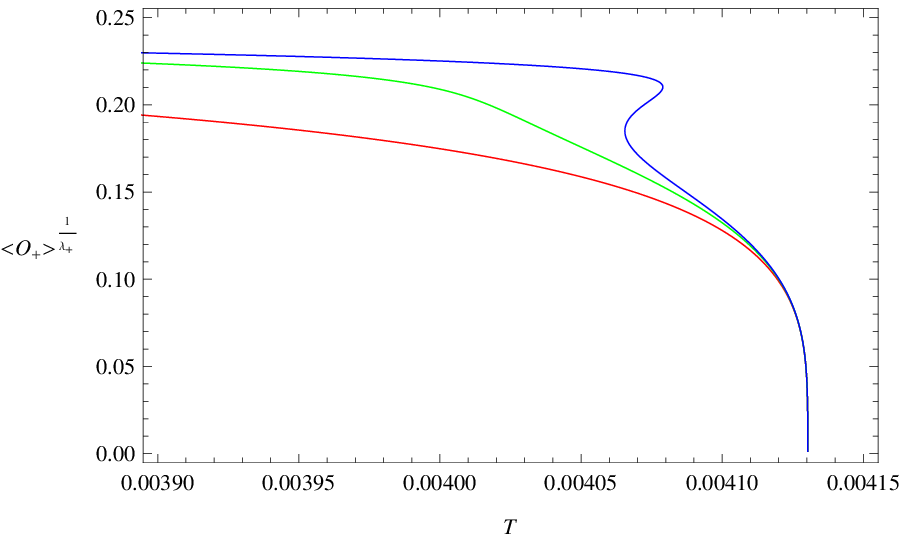}\
\caption{\label{EEntropySoliton} (Color online) The condensation of the scalar operators. The left panel corresponds to the case of $\gamma=0.1$, $m^{2}=-2$
, $T_{c}=0.0149$ and $\zeta$ varies as $\zeta=0$ (Red), $\zeta=0.01$ (Green), $\zeta=0.02$ (Blue). The right one shows the case $\gamma=0.1$, $m^{2}=-2$
, $T_{c}=0.0041$ and $\zeta$ varies as $\zeta=0$ (Red), $\zeta=0.003$ (Green), $\zeta=0.004$ (Blue).}
\end{figure}

In order to study the stability of the states,
we show the free energy of the systems corresponding to different states with $\gamma=0.1$, $m^{2}=-2$ and $\zeta=0$ in Fig. 8.
We find that the higher state corresponds to larger free energy.
So we can label
superconductor solutions as the first, the second and the third energy states through various increasing grand canonical free energy.
The solution in (a) of Fig. 1, Fig. 2 and Fig. 4 corresponding to the bottom red line in Fig. 8 has the lowest free energy and is
thus the stable phase, whereas the solutions with higher free energy are unstable.

\begin{figure}[h]
\centering
\includegraphics[width=220pt]{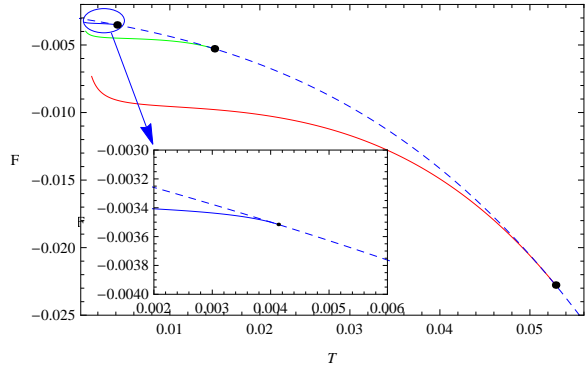}\
\caption{\label{EEntropySoliton} (Color online) The picture is for the behaviors of free energy with respect to the temperature with $\gamma=0.1$, $m^{2}=-2$ and $\zeta=0$. The curves from bottom to top correspond to the first (red), the second (green) and the third (blue) superconducting states. We have choose the chemical potential $\mu=1$ in both states.
The blue dashed line shows the free energy of the pure AdS black hole.
The black solid points correspond to the critical phase transition point between normal and superconductor phase.}
\end{figure}

\section{Conclusions}

We investigated a general class of holographic superconductors via
St\"{u}ckelberg mechanism in the background of
AdS black hole.
We obtained richer structure in the metal/superconductor phase transitions.
We observed that the model parameter
coupled with the scalar mass and backreaction
can determine the order of
phase transitions.
We found two threshold values $\tilde{\zeta}$ and $\bar{\zeta}$ for each pairs of scalar mass $m^{2}$ and backreaction parameter $\gamma$.
When the model parameter satisfying $\zeta\leqslant\tilde{\zeta}$, there is only the second order phase transition at $T_{c}$.
If $\tilde{\zeta}<\zeta<\bar{\zeta}$, the phase transition at $T_{c}$ is still the second order. However, an additional first order phase transition
appears in the superconducting phase. When $\zeta\geqslant\bar{\zeta}$, it is the typical first order phase transition at the phase transition
point $T_{c}$.
We also tried to disclose the properties of the phase transitions by analyzing the
entanglement entropy of the metal/suoerconductor system.
We argued that the entanglement entropy serves as a good probe
to the order of the phase transitions and
the jump of the entanglement entropy would be a quite general feature for
the first order phase transition in the AdS black hole background.
In addition, we examined effects of the scalar mass and backreaction on the scalar condensation.
We found that the less negative
mass and stronger backreaction make all types of first order phase transitions easier to happen. Furthermore, we arrived at a relation $\overline{\zeta}\thickapprox 2 \widetilde{\zeta}$ between
the threshold parameters.
With the shooting method, we obtained various
superconductor solutions corresponding to different energy states.
At last, we disclosed the stability of various energy states with the behaviors of the free energy.
We concluded that the usually studied holographic superconductor solutions corresponding to the lowest energy state is stable, whereas the higher energy state
is unstable.

\begin{acknowledgments}

We would like to thank Professor Bin Wang for his helpful discussions and suggestions on this topic.
We also want to thank the anonymous referee for the constructive suggestions to improve the manuscript.
This work was supported in part
by the National Natural Science Foundation of China under Grant No. 11305097, the
education department of Shaanxi province of China under Grant No. 2013JK0616 and the Foundation of Shaaxi University of Technology
of China under Grant No. SLGQD13-23. This work was also supported by the China postdoctoral Science Foundation under Grant No. 2013M531163.

\end{acknowledgments}

\end{document}